# Advection-enhanced kinetics in microtiter plates for improved surface assay quantitation and multiplexing capabilities

Iago Pereiro[a] §, Anna Fomitcheva Khartchenko[a] §, Robert D. Lovchik[a] and Govind V. Kaigala*[a]

§ Equal contribution

[a]Dr. I. Pereiro, Dr. A. Fomitcheva Khartchenko, Dr. R.D. Lovchik, Dr. G.V. Kaigala

IBM Research – Europe, Säumerstrasse 4, Rüschlikon CH-8803, Switzerland

E-mail: gov@zurich.ibm.com

**Abstract:** Surface assays such as ELISA are pervasive in clinics and research and predominantly standardized in microtiter plates (MTP). MTPs provide many advantages but are often detrimental to surface assay efficiency due to inherent mass transport limitations. Microscale flows can overcome these and largely improve assay kinetics. However, the disruptive nature of microfluidics with existing labware and protocols has narrowed its transformative potential. We present WellProbe, a novel microfluidic concept compatible with MTPs. With it, we show that immunoassays become more sensitive at low concentrations (up to 9× signal improvement in 12x less time), richer in information with 3-4 different kinetic conditions, and can be used to estimate kinetic parameters, minimize washing steps and non-specific binding, and identify compromised results. We further multiplex single-well assays combining WellProbe's kinetic regions with tailored microarrays. Finally, we demonstrate our system in a context of immunoglobulin subclass evaluation, increasingly regarded as clinically relevant.

**Introduction**

Solid phase assays to detect and quantify analytes are essential in diagnostics and screening[1,2], drug discovery[3], environmental monitoring[4] and quality control[5,6], among others. The use of a solid phase to immobilize ligands or analytes enables separation steps, while a flat interface further facilitates optical reading and multiplexing. As a result, many bioassays today rely on solid surfaces, here referred to as "surface assays", including almost all immunoassays.[7] Well-known examples of surface assays are enzyme-linked immunosorbent assays (ELISA), immunohistochemistry, immunofluorescence and microarrays. In research, just these four assay types account for more than 100k new publications per year (results of a keyword search for these assay names in Scopus for the year 2019). In clinical practice, ELISA alone has become the gold standard for diagnosing many viral, bacterial and parasitic diseases, including HIV, Hepatitis B and C, influenza, paratuberculosis and Lyme disease.[1]

Most surface assays are carried out in multi-well microtiter plates (MTPs), particularly in the 96-well format, as they allow multiple parallel tests in independent wells with minimal cross-contamination. It is noteworthy that MTPs have barely changed since they started being injection-molded in the 1980s. Their standardization in the early 2000s[8] and the emergence of dedicated equipment in the form of dispensers, washers, readers and automated on-bench instrumentation has further helped consolidate their use. And yet, this consolidation has occurred in spite of notable intrinsic drawbacks for surface assays. Notably, incubation steps typically require 1-8 hours or overnight to enable good sensitivity while compensating this with relatively high analyte concentrations, often making reagent/sample use inefficient. This is further compromised by evaporation, particularly for high yield plates with many (>1000) wells.

These drawbacks are associated with inefficient mass transport within the assay, or more specifically, a slow diffusion process of the analytes in the liquid towards the reaction surface. This is particularly prominent in the case of biomolecular assays, as proteins,[9] long DNA/RNA strands[10] and nano/micro-particles are often associated with low diffusivities. In the duration of the assay, the zone of the liquid adjacent to the reaction surface can become depleted in analyte. As this zone grows in size the reaction rate diminishes, significantly increasing the required incubation times. To overcome this, two approaches are possible: (i) using a higher analyte concentration in the liquid to accelerate diffusion by increasing the concentration gradients, albeit consuming more analyte, or (ii) generating flows (advection) within the liquid to constantly bring fresh analyte close to the surface, thereby reducing the thickness of the depletion zones. Overcoming transport limitations in the latter way can lead to gains in the reaction rate of several orders of magnitude.[11] Therefore, to obtain advection in surface assays, MTPs are often placed on orbital shakers to slosh the liquid in the wells. However, the obtained flows are of limited intensity and mostly present in the liquid bulk far from the surface. Thus, signal gains are limited and often coupled to non-uniformity and low reproducibility, resulting in suboptimal quantitation accuracy.[12]

In contrast, a more precise control of flows can be obtained at the microscale, where hydrodynamics are dominated by deterministic laminar regimes. By integrating and miniaturizing many analytical functions in small devices, microfluidics has enabled notable applications in genomics,[13] cancer diagnostics,[14] single cell studies[15] or material discovery[16], to name a few. In the case of surface assays, microfluidics can significantly enhance kinetics and largely improve the time to results and reproducibility.[17–19]

However, few commercially available systems exist to enhance surface assays with microfluidics, despite three decades of development. A key reason for this is that non-standardized, custom-designed and non-transferable microfluidic systems offer little versatility and compatibility with existing and well-established surface assay labware, protocols and practices. To try to overcome this limitation, a few attempts have been made to create microfluidic devices with an external geometry resembling the MTP format. The primary aim was to make them compatible with common plate readers and dispensers. For example, Witek et al. presented a nucleic acid purification device, fabricated using hot embossing, containing 96 purification beds consisting of individual chambers with microposts.[20] Hou et al. reported a PDMS microfluidic chip with channels to dilute the sample at different concentrations and let it react in chambers arranged as wells of a 48 well plate.[21] To avoid the need for external fluidic pumping, Kai et al. presented a device with an external geometry identical to a 96-well plate, incorporating a spiral microfluidic channel at the bottom of each well to drive the dispensed liquid towards an absorbent pad.[22] Similarly, Sanjay et al. presented a PMMA/paper hybrid device in which the dispensed reagents in the wells were transferred onto a paper-based assay.[23] In other cases, the microtiter plate format was miniaturized, to benefit from its parallelization in a portable format.[24] All these systems have in common that the resulting devices need to be customized, are still of limited compatibility with standard equipment and are not compatible with commercially available and well-established MTPs. Recently, an MTP-compatible cover comprising microfluidic channels was reported, allowing liquid exchange in MTP wells.[25] Such a system is useful for medium renewal in cell cultures, but unfortunately of limited use for biomolecular surface assays, as it does not enable controlled advection at the well surface.

Here, we present a new microfluidic concept to address standard microtiter plates while enhancing surface assays with accurate advection conditions. We call this concept "WellProbe". Performing MTP-based surface assays with WellProbe provides the following advantages: (i) it generates flow-enhanced kinetics, which considerably reduce bioassay time and increase sensitivity, (ii) it provides several local kinetic conditions within one well, expanding the linear quantitation range of the assay, (iii) it provides kinetics-based information with a single test, which can serve to identify saturated conditions, identify artefacts and false positives or false negatives, and strengthen quantitation accuracy, (iv) it enables rapid liquid switching for multi-step protocols while avoiding cross-contamination and making intermediate washing steps unnecessary, (v) coupled with arrays of radially distributed spots, it allows high-level multiplexing in single-well tests, and, (vi) it is compatible with commercially available MTP-based kits for surface assay applications, as well as the instrumentation for reading results or dispensing additional liquids. We demonstrate the functionality of our system in the context of IgG binding, a step virtually compulsory in all ELISA tests, and illustrate its application for subclass IgG (IgGSc) testing. Measuring the four subclasses of IgG (IgG1, IgG2, IgG3 and IgG4) can provide a more complete view of the function of the humoral immune system than the more common testing of immunoglobulin classes IgG, IgA and IgM[26]. However, the concentration range of each subclass in adult human plasma differs strongly (IgG: 767-1590 mg/dL IgG2: 171-632 mg/dL, IgG4: 2,4-121 mg/dL).[27] Such variations can represent a challenge for accurate quantitation with standard ELISA assays, where the quantifiable range is obtained by blind sample dilution and pre-defined assay timing for each analyte. We thus regard IgGSc as a particularly pertinent target for more accurate and multiplexed testing, which could contribute to both increase our understanding of IgGSc deficiency and make its testing more readily available in the clinics.

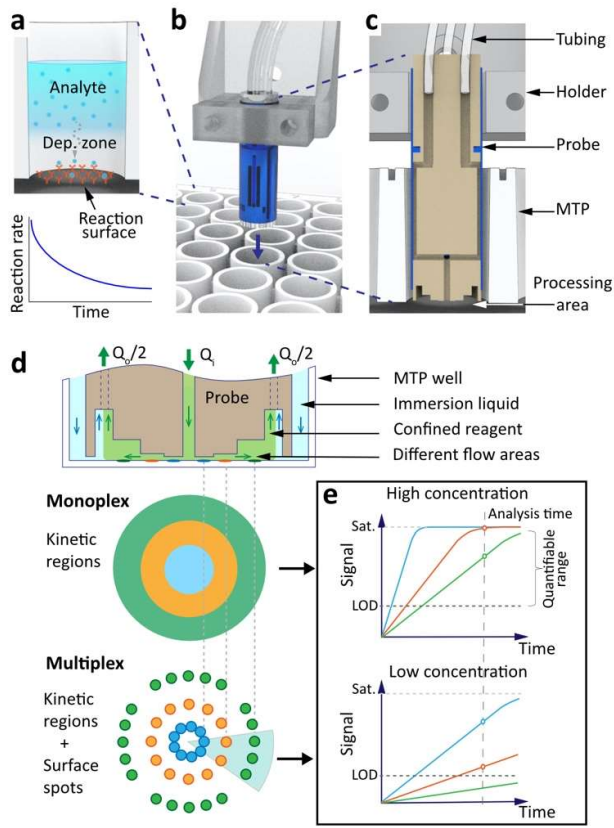

*Figure 1:* Working principle of WellProbe. (a) Common case of a surface assay in a well of a static MTP, with creation of a depletion zone that leads to reaction rate decay; (b) WellProbe positioned over an MTP well and (c) cross section of WellProbe within a well; (d) appropriate injection/aspiration flow-rates result in a radial flow that creates different kinetic regions on the surface, which can either be used for monoplex assays or multiplexed assays with spots; (e) the kinetic regions provide an expanded quantifiable range.

**Results and Discussion**

**Working principle**

Figure 1a illustrates a typical surface assay in a well of a static MTP. The reaction surface consumes analyte in the adjacent liquid, creating a depletion zone of low analyte concentration that gradually increases in size. The remaining analyte in the liquid bulk needs to traverse this zone by diffusion to reach the reaction surface, making the reaction rate decay with time. Figures 1b-c show the concept of WellProbe, which consists of a cylindrical microfluidic device that can enter individual MTP wells to address the bottom surface. In contrast to previous open microfluidic probes that localize reagents in microscale regions and need to scan the surface to address larger areas[28–30], WellProbe creates instant circular flow confinements in the mm- and cm-scale, covering most of the well bottom. At its tip, a central injection opening is surrounded by circular flat areas, or mesas, of increasing height (Figure 1d). The liquid exits the injection opening radially and exhibits a decreasing flow velocity regime and shear rate below each section of the mesa. In transport-limited surface reactions - common for biomolecules e.g. relatively large proteins such as antibodies, enzymes, hormones or globular proteins - an increased shear typically translates into reduced depletion layers and higher binding rates.[31] As illustrated in Figure 1d, under each mesa a region of particular kinetic conditions is created where analytes can bind to ligands on a uniform surface or in discrete spots. Thus, assay dynamics are locally unique, with higher binding rates in the inner regions (Figure 1e). Two working modes are possible: (i) monoplex, with standard MTP wells containing one specific ligand and (ii) multiplex, combining the kinetic regions with a tailored-distribution of surface spots. In all cases, the system condenses several traditional assay tests into one single test, provides a larger quantitation range and kinetics-based information and shows when the signal is saturated or has been compromised.

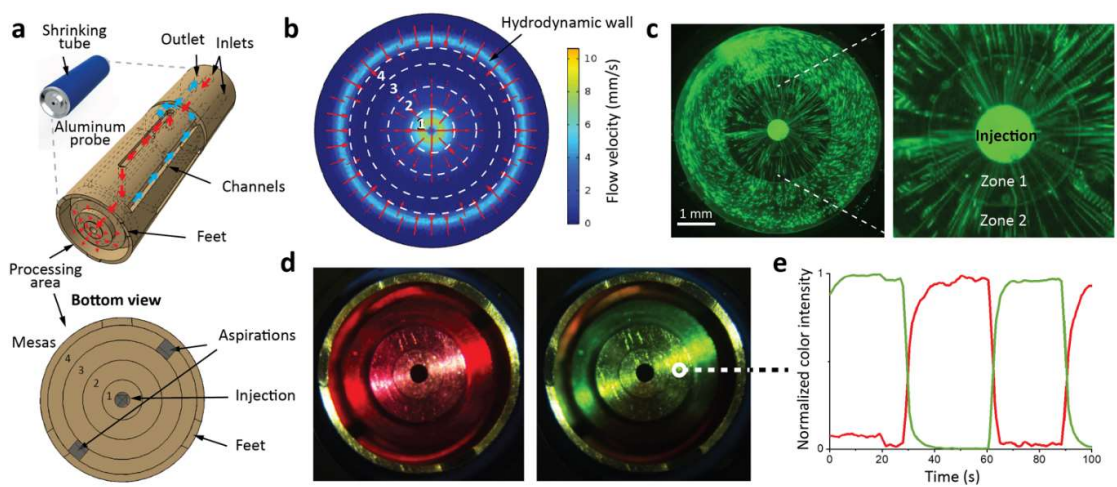

*Figure 2. WellProbe radial hydrodynamics and liquid exchange for sequential chemistry. (a) Flow paths within WellProbe and bottom view with mesas; (b) FEM hydrodynamic simulation of the flow confinement area; (c) experimental streamlines with fluorescent particles; (d, e) intermittent switching of colored solutions.*

**WellProbe-generated surface hydrodynamics**

Figure 2a illustrates the internal structure of WellProbe. The aluminum body (diameter 5.5 mm, length 20 mm, see SI-S6.2 for details) contains an internal channel network, created by sealing side-drilled channels and surface grooves with a polymer thermal shrinking tube, thus becoming closed channels. In the configuration shown, WellProbe contains four aperture inlets for sequential liquid switching between up to four reagents and a single outlet aperture. The channels originating at the aperture inlets join 5 mm upstream of the tip aperture (the red arrows in Figure 2a follow one path). The distance between WellProbe and the well bottom is defined by peripheral feet (50 μm). After flowing over the processing area below the three mesas surrounding the central aperture (with increasing heights of 50, 100 and 500 µm), the reagent is collected by two symmetrically opposed apertures in a fourth external mesa of 1.2 mm height. By collecting the liquid using a 1 mm deep recession and two aspiration apertures, we ensure that the aspiration at the well bottom is constant and the resulting streamlines are perfectly radial with symmetrical shear stress in all directions (Figures 2b-c and SI-S2). This recession further acts as a bubble trap[32], collecting any undesired air coming through the aperture, increasing system robustness and reproducibility (see SI-S6.1). Using an aspiration flow-rate Qa slightly higher than the injection flow-rate Qi (Qi = 40 µL/min and Qa = 45 µL/min in Figure 2b) results in a 5 mm wide circular confinement of the injected reagent, 80% of the total well diameter. This confinement is entirely surrounded by a hydrodynamic wall of external immersion liquid (buffer) entering WellProbe through a circular groove near the tip edge, which sets the boundary of the flow confinement and prevents cross-contamination (Figure 2b). The channels with the liquid aspirated through the two apertures (reagent plus external immersion liquid) join within WellProbe and lead to a single outlet aperture (blue arrows in Figure 2a). The inlet channels join close to the tip of WellProbe, and if these channels are purged before the experiment, the reagents in the confinement can be switched very quickly by using upstream valves for each inlet. In Figures 2d and e, two colored reagents are sequentially switched with a transition time of less than 5 s at 40 µL/min. This allows for efficient multi-step assays.

**Advection-enhanced kinetics enabled by WellProbe**

A key feature of WellProbe is that it allows to enhance the kinetics of surface bioassays by creating high and accurate shear rate (and shear stress) conditions at the well bottom, which reduces and stabilizes the depletion layers. This shifts the kinetics from being transport-limited to reaction-limited. The shear rate ($\dot{\gamma}$) at any point of the well bottom below WellProbe is calculated as the local derivative of the flow profile between two infinite parallel plates, a solution of the Navier Stokes equations (FEM simulations in Figure 3a):

$$\dot{\gamma} = \frac{3Q}{\pi R H^2} \quad (1)$$

where Q is the injection flow-rate, R is the radius or horizontal distance between the injection and the position of interest and H is the vertical distance between the well bottom and WellProbe. The shear rate obtained under mesa 1 is 500× higher than under mesa 3 (e.g. for Q = 40 µL/min, $\dot{\gamma}\_1$= 765 s-1 and $\dot{\gamma}\_3$= 1.5 s-1, calculated at the middle diameter of each mesa). The sharp transition between mesa heights ensures that shear rate conditions under contiguous mesas have a well defined boundary.

The value of the shear rate indicates the magnitude of the advective flows close to the surface. Higher values result in improved transport and thus smaller depletion layers. The ratio between the advective and diffusive transport rates, or Peclet number, is high (Pe >> 1) in the working flow-rate regimes of WellProbe (> 1 µL/min). Therefore, using Eq. 1, we estimated the thickness of the depletion layer (δ) on the well bottom as the approximate distance from the bottom at which a particle requires equal time to reach the bottom by diffusion or escape the region by advection.[17] For this, we took into account the transport under each mesa more central than the mesa of interest n (e.g. mesas 1 and 2 when measuring mesa 3, see details in SI-S3):

$$\delta = \left( D \frac{\pi}{3Q} \sum_{i=1}^{n} \frac{H_n^3}{H_i} (R_i^2 - R_{i-1}^2) \right)^{\frac{1}{3}} \quad (2)$$

where Ri is the external radius of each mesa considered, Hi its height and D is the diffusivity of the molecule being transported. Thus, e.g. at Q = 40 µL/min and a common IgG diffusivity of D = 6.5 x $10^{-5}$ mm2/s[33,34] we estimate the depletion layer thickness under mesas 1-3 to be approximately 2.3, 11 and 66 μm, respectively. Additionally, we performed 2D radially symmetric FEM simulations of the expected concentration profiles, as shown in Figure 3c. For the same Q and D, the obtained depletion layer thicknesses were 2.2, 13 and 84 μm under mesas 1, 2 and 3, respectively (Figure 3c), which is in good agreement with the analytical values.

If we consider the case of transport-limited kinetics, so that we assume that all analytes reaching the surface also bind to it, and further neglecting the saturation of the binding sites on the surface, we can approximate the local binding rate ($\dot{b}$) as being equal to the rate of diffusion through the depletion layer:

$$\dot{b} \approx D \frac{c_0}{\delta} \tag{3}$$

where $c_0$ is the concentration of the molecule of interest in the injected reagent. The potential difference in binding kinetics between mesas 1 and 3 (more precisely the mesa's central radius $R_1$ = 0.33 mm, $R_2$ = 1 mm, $R_3$ = 1,67 mm) is $\dot{b}_1/\dot{b}_3$ ~30 (precisely 29.2) and between mesas 1 and 2 $\dot{b}_1/\dot{b}_2$ ~5 (precisely 4.8). The resulting binding kinetics can thus cover more than one order of magnitude with three points similarly spaced on a logarithmic scale.

**Single-well antibody binding with 3 simultaneous kinetic conditions**

As seen in Figure 3b, the resulting kinetics under each mesa translate into proportional signals in an IgG – anti-IgG reaction. To observe the temporal kinetic behavior (Figure 3d, experimental details explained in SI-S1 and supporting chemical characterization in SI-S5), we injected two fluorescent-IgG concentrations (at 1 and 10 μg/mL) for up to 6 minutes (incubation time) at 40 μL/min. As observed, the inner regions consistently exhibited a higher signal intensity than the outer ones, except when reaching saturation. At incubation times 3 and 6 minutes and 10 μg/mL, similar signal levels in mesas 1&2 indicate that saturation is being reached. As expected, in conditions far from saturation and well above LOD levels (e.g. 6 min with 1 μg/mL) the signals from mesas 1 and 3 differ by approximately one order of magnitude. Depending on the incubation times tested, quantification was possible in either the areas with fast kinetics (mesas 1 & 2) for challenging low concentrations or within the slow kinetic areas (mesa 3) when the signal approached saturation.

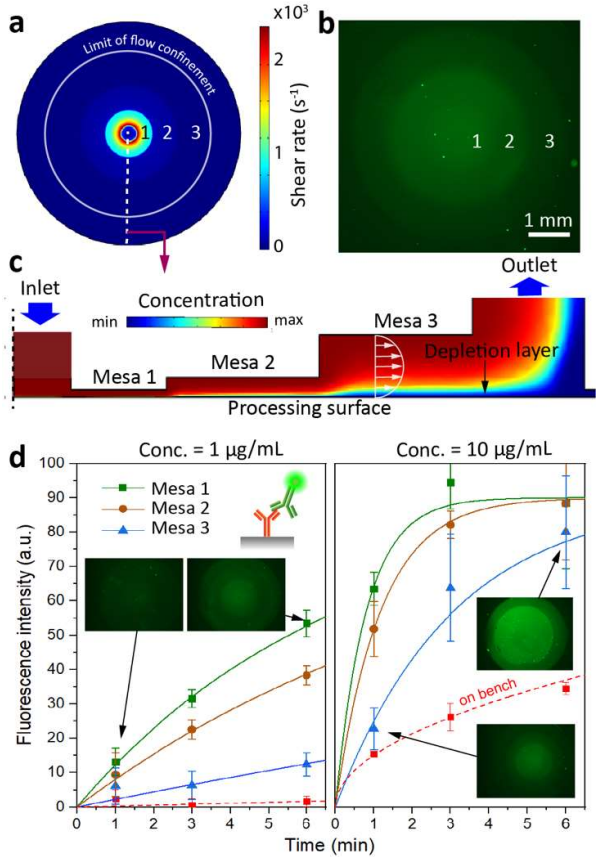

*Figure 3:* Shear rate-dependent IgG binding dynamics. (a) Simulation of the shear rate at the well bottom with indicated position of mesas; (b) Fluorescent signal corresponding to 3 mesas after one test (40 µL/min, 10 µg/mL, 1 min); (c) simulation of analyte concentration profile and depletion layers in section view of confinement; (d) binding curves of an Ab-Ab system for different incubation times and antibody concentrations at an injection flow-rate of 40 µL/min (n = 3).

From such curves, obtained from a few wells, it is possible to estimate the values of some kinetic constants. In SI-S2.3, we use the results from Figure 3d at 10 µg/mL to obtain fitting curves with a finite element model based on first-order surface kinetics. Considering the value of D previously mentioned and a typical dissociation constant $k_{off} = 1 \times 10^{-3}$ s$^{-1}$, we estimated that the concentration of binding sites on the well surface is approximately $b_m = 3 \times 10^4$ mol/cm$^2$ and the affinity constant $k_{on} = 4 \times 10^5$ L/mol·s, a value comparable to those reported for Ab-Ab interactions in the literature[35].

Figure 3d additionally shows dashed red curves obtained from a classical on-bench assay, in which all conditions are kept identical expect that transport is purely diffusion based (no WellProbe used). As can be seen, even zone 3, the zone with the lowest flow velocities, exhibits significantly higher signal intensities than on-bench results. To better quantify the benefits of using the WellProbe system, we can estimate the gains in assay time (ε) with the following expression:

$$\varepsilon = \frac{\tau_{diffusion}}{\tau_{reaction}} \approx \frac{2 b_m^2 K^2 k_{off}}{3D(1 + Kc_0)} \quad (4)$$

where $K = k_{on}/k_{off}$ (the ratio of the association rate to the dissociation rate), $b_m$ is the density of binding sites on the surface, $\tau_{diffusion}$ and $\tau_{reaction}$ are the approximate times to reach signal saturation (kinetic equilibrium) by relying purely on transport-limited diffusion (experiment on-bench) or using an ideal system with negligible transport limitations (an efficient use of WellProbe, see SI-S3 for how we obtain this expression). This equation particularly reveals how the expected gains increase with decreasing analyte concentrations. For example, with our Ab model system, we would expect that for 10 µg/mL and 1 µg/mL, WellProbe could reach kinetic equilibrium at 12× (2 instead of 22 min) and 93× (13 min instead of 21 h) shorter assay times respectively, compared to on-bench protocols. Note that equilibrium does not need to be reached for assays to provide quantitation. In SI-S5.1 we performed a calibration of our model system, confirming that for lower concentrations (<1 µg/mL) WellProbe provides statistically better signal than equivalent on bench protocols of 20× the incubation time (3 min instead of 1 h) and an order of magnitude lower LOD for an equivalent assay time. We further confirmed this by performing ELISA tests at a lower concentration of 100 ng/mL (SI-S5.2), showing ~9× better signal with WellProbe in 5 min than on bench for 1 h of primary antibody incubation. Additionally, non-specific binding is also significantly reduced due to the constant flow-rate (SI-S5.3). In summary, WellProbe provides an increased signal level at low concentrations and improves the limit of detection (LOD) as compared to standard assays.

**Hydrodynamic reconfigurability expands the quantitation range**

As with any assay, sometimes quantitation may be compromised. WellProbe provides the advantage of revealing such situations by analyzing the obtained signal patterns in mesas 1-3 (see SI-S4). In one scenario, samples with a high concentration of analytes can lead to signal saturation (see SI-S5.4 for upper range of model system). To further increase the upper range of measurable analyte concentrations, it is possible to use the dynamic reconfigurability of the flow confinement to address an additional zone 4 of the surface for a different incubation time. As shown in Figure 4a, by increasing the ratio of outlet aspiration ($Q_o$) to inlet injection ($Q_i$), the area containing the confined reagent reduces its width, covering mesas 1, 2 and partially 3. In contrast, a ratio close to 1 (similar injection to aspiration) tends to cover the entire area below WellProbe (Fig. 4b, covering mesas 1, 2, 3 and 4). In Figure 4c we demonstrate the use of this dynamic feature to measure an IgG concentration of 50 µg/mL. While mesas 1, 2 and 3 were exposed to the analyte for 3 minutes, the area below mesa 4 was only exposed for the last 10% of this time (18 s). While these experimental conditions lead to saturation levels being reached in mesas 1-3, mesa 4 exhibits signal within the measurable range. Thus, coupling this dynamic reconfigurability to the system of fixed mesas allows to extend the quantitation range of surface assays by at least two orders of magnitude.

**Coupling kinetic zones and spots to multiplex for IgG subclass quantitation**

Lastly, we performed single-well multiplexing by coupling WellProbe with custom-made radial arrays of spots deposited by inkjet printing in a standard MTP (Figure 5a). Microarrays in MTPs offer the advantage of being easier to automate and adapt to biochemical protocols[36], but the concept has been little explored primarily due to diffusion limitations[37]. Here, we inkjet-printed arrays consisting of three concentric rings of spots, each ring containing 9 spots with 3 replicates for each tested IgG, for a 3-plex. The spatial distribution of each ring of spots coincides with one of the 3 mesas. The first circle of spots is located closer to the inlet mesa radius (at R = 0.33 mm) to leverage the locally higher shear rate (see SI-S6.3 for information on spot configuration).

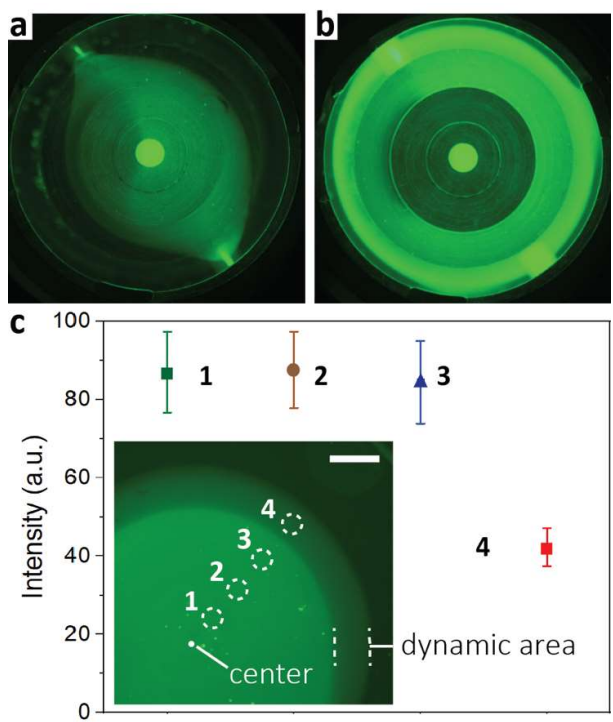

*Figure 4:* Expanded quantitation range with confinement reconfigurability (a) Shrunk flow confinement with an injection:aspiration ratio of 1:3 ($Q_i$ = 40 µL/min, $Q_o$ = 120 µL/min). (b) Flow confinement covering the entire tip of WellProbe with a ratio close to 1 ($Q_i$ = 40 µL/min, $Q_o$ = 41 µL/min). (c) Results of an assay with IgG at 50 µg/mL (n = 3) for a total of 3 minutes under mesas 1-3 and the *last 18 seconds additionally under mesa 4 (Qi = 40 μL/min, initial Qo = 41 μL/min and final Qo = 100 μL/min); the insert shows a fluorescence image of the obtained assay signal.*

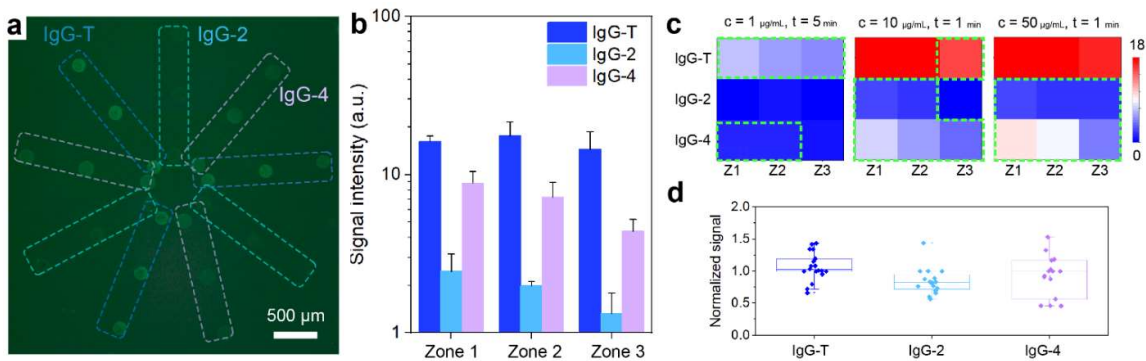

*Figure 5: Multiplexing combining kinetic areas and circular distribution of spots. (a) Radial array of spots after an assay with WellProbe, containing triplicate spots per zone and specificity for all IgG subtypes and subtypes 2 and 4. (b) Results obtained from a single well/array (sample was human IgG protein at 10 μg/mL, incubation time 1 min), with standard deviations obtained from in-well spot triplicate. (c) Heatmaps of individual arrays indicating average spot intensity for each spot type and zone at different dilutions of same sample or incubation time: dotted green line indicates measurable spots, neither in saturation nor close to LOD. (d) Results from all measurable averaged spots obtained from concentrations 1, 10 and 50 μg/mL and incubation times of primary antibody of 1 and 5 min (2 wells per condition): all results normalized to results from well at 10 μg/mL and 1 min.*

We applied this multiplexing to IgG subclass testing. For certain subclasses, notably IgG2 and IgG4, concentration values below the normal healthy range are correlated with an impaired response to infections in the respiratory tract[38], allergic asthma, rhinitis and autoimmune conditions.[26] In the case of IgG4, elevated levels are associated with autoimmune pancreatitis, aspirin-exacerbated respiratory disease, nasal polyps, eosinophilia and celiac disease.[39] In spite of their increasing clinical relevance, a standardization of IgGSc assays is still lacking.[27] In this work, the spots contain antibodies specific to the capture of total IgG and subclasses IgG2 and IgG4. As sample, we used purified human IgG containing a representative mix of IgG subclasses in human plasma. This sample was injected with two different incubation times (1 and 5 min) at 40 μL/min and three dilutions in PBS: 1, 10 and 50 μg/mL. This was followed by liquid switching to inject the secondary fluorescent antibody for 5 min (enough to ensure saturation and avoid zone-dependent binding). A washing step between reagents was deemed unnecessary as the constant flow ensures that non-bound species are removed from the spots. Figure 5b shows representative results extracted from spot signals of a single well/array (c = 1 μg/mL, t = 1 min). A descending trend is observed between zones 1 and 3 for IgG2 and IgG4, indicative of spot signal in the measurable range and good assay quality. The approximately constant values for IgG-T evidence that saturation has been reached. This is easier to visualize in the form of heatmaps, as depicted in Figure 5c, where each square represents the average spot intensity of spots of the same zone and IgG subtype. When the signal level is less than 3 times the noise and the square color shows a saturation plateau (values of zones 1- 2 or 1-3 were not statistically different), those values are considered inadequate for quantitation. To estimate the quantitation accuracy of the system, we used all selected values of Fig. 5c (marked in green) and, assuming a linear behavior, normalized with respect to the results of one condition (c = 10 μg/mL, t = 1 min) and multiplied by a factor corresponding to their difference in dilution or time: e.g., c = 1 µg/mL, t = 5 would be multiplied by 10/5=2. The resulting equivalent values in Figure 5d show that in spite of the large range studied, all estimated values fall within a 50% confidence interval for all IgG subtypes. Thus, a single test can offer multiplexed results, quality assessment and an indication of the measurable range of quantitation, while enabling replicates within the same experimental conditions.

**Conclusion**

Compared to on-bench diffusion-based tests, we have shown that signal enhancements of up to an order of magnitude are possible while supplying 3-4 times more information per well in the monoplex configuration. Additionally, the 3-mesa design shown here can be expanded to more mesas if an increased resolution in shear rate conditions is desired.

Higher flow-rates result in enhanced kinetics until transport ceases to be the limiting factor. For example, the 40 µL/min used for the experimental results in Figure 3 result in a Damköhler number Da = 0.4, very close to ideal reaction-limited kinetics. In this case, a higher flow-rate would lead to unnecessary waste of sample. Eq. 2 and 3 show that the binding rate is approximately proportional to the cube root of the flow-rate. Thus, reducing the flow-rate by half only results in a 20% loss in signal, something the user can leverage to optimize signal and reagent use for precious samples. That said, the results illustrated in this work used sample volumes of 60-250 µL, comparable or smaller than typical volumes in standard ELISA assays, with all the added benefits. Aspirating the sample through WellProbe prior to injection can ensure that only the necessary volume is used, thereby avoiding partial loss of the sample in internal channels.

The WellProbe concept is adaptable to MTPs of any size and shape. The specific probe shown here is reusable and inert to biological substrates due to the aluminum oxide layer that naturally covers its structure, but similar results could be expected with plastic, titanium or steel. For sequential chemistry, the design shown here provides four inlets. While intermediate washing steps are less needed with advection, some protocols still require a high number of reagents. Although a higher number of inlets is possible within the shown device, external junctions containing multiple inlets could be placed upstream of WellProbe to accommodate complex protocols requiring many reagents. Additionally, while WellProbe has the benefit of providing the results of many wells in a single well-test, increased throughput can be obtained with multiple WellProbes working in parallel, if needed. Furthermore, the same concept could be envisioned in integrated and disposable mass-manufacturable devices. In this case we foresee that pipetting robots could be used for direct liquid injection into the inlets, to facilitate liquid handling automation.

The predominant use of MTPs in life science settings is due to so far unchallenged benefits in ease of use, accessibility, parallelization, low cross-contamination and compatibility with a myriad of now standard laboratory readers, washers or pipettors. Thirty years of microfluidic alternatives have had limited impact on this continued use of MTPs, which have barely been replaced for such ubiquitous bioassays as ELISA. Instead, new robotic and automated platforms have been created to operate and automate MTP-based assay workflows. Here, we combine the benefits of microfluidics and the intrinsic advantages of existing MTPs for the first time, thereby increasing their multiplexing capability, sensitivity and accuracy. Undeniably, the current demand for better quantitation, data acquisition and data quality will continue in the next decades to feed the new needs of the biological and biomedical communities. We see WellProbe as a key step to adapt many of the current workflows to the requirements of next generation diagnostics and quantitative biology, presenting a low implementation barrier at the same time.

**Acknowledgements**

We thank A. Zulji for device fabrication and L. Petrini for detailed discussions. We acknowledge funding by ERC-PoC CellProbe (842790) and SNF Spark program (CRSK-2_190877). We thank Dr. E. Delamarche and Dr. H. Riel for continuous support.

**Keywords:** surface assays, open-space microfluidics, microtiter plate, multiplexing, advection, mass transport